\documentclass{pkas}
\def\year{2016} 
\def\month{March} 
\def\volume{00} 
\def\issue{0} 
\def\beginpage{1} 
\def\endpage{2} 
\def\received{---} 
\def\accepted{---} 
\date{Received \received ; accepted \accepted}
\usepackage{color}

\title{
A quality check of the $AKARI$ mid-infrared all-sky diffuse map toward the massive star-forming regions NGC 6334 and NGC 6357
}

\author[1]{Hidetoshi Sano}
\author[1]{Tomoya Amatsutsu}
\author[1]{Toru Kondo}
\author[1]{Keichiro Nakamichi}
\author[1]{Mitsuyoshi Yamagishi}
\author[1]{Daisuke Ishihara}
\author[1]{Shinki Oyabu}
\author[1]{Hidehiro Kaneda}
\author[1]{Kengo Tachihara}
\author[1]{Yasuo Fukui}

\affil[1]{Department of Physics, Nagoya University, Furo-cho, Chikusa-ku, Nagoya 464-8601, Japan; \email{sano@u.phys.nagoya-u.ac.jp}}

\def\corrauthor{
Hidetoshi SANO
}

\def\runningauthor{
Sano et al.
}

\def\runningtitle{
}

\def\keywords{
infrared: surveys -- massive star-forming regions: NGC 6334 and NGC 6357
}

\def\abstracttext{
We present a comparative study of CO and polycyclic aromatic hydrocarbon (PAH) emission toward a region including the massive star-forming regions of NGC 6334 and NGC 6357. We use the NANTEN $^{12}$CO($J$=1--0) data and the $AKARI$ 9 $\mu$m All-Sky diffuse map in order to evaluate the calibration accuracy of the $AKARI$ data. We confirm that the overall CO distribution shows a good spatial correspondence with the PAH emission, and their intensities exhibit a good power-law correlation with a spatial resolution down to 4$\rq{}$ over the region of 10$^\circ$$\times$10$^\circ$. We also reveal poorer correlation for small scale structures between the two quantities toward NGC 6357, due to strong UV radiation from local sources. Larger scatter in the correlation toward NGC 6357 indicates higher ionization degree and/or PAH excitation than that of NGC 6334.}

\begin{document}
\pkashead 


\section{INTRODUCTION}
$AKARI$ ($ASTRO$-$F$), the infrared astronomical satellite operated by Japan Aerospace Exploration Agency (JAXA), is producing a new vision of our universe in the infrared wavelength. Especially, the All-Sky survey by using the Infrared Camera \citep[IRC;][]{2007PASJ...59S.401O} revealed a whole sky distribution of the emission from polycyclic aromatic hydrocarbons (PAHs), which provides important information to understand the origin of organic matter.

We are preparing for the public release of the $AKARI$ mid-infrared All-Sky diffuse map in 9 $\mu$m $\&$ 18 $\mu$m containing the PAH emission. The basic and advanced calibrations including removal or correction of various artifacts (Amatsutsu et al., 2016 in prep.) are almost completed. Thus, the calibration accuracy needs to be evaluated in a large scale. In this paper, we investigate the correlation between the $AKARI$ 9 $\mu$m emission and the NANTEN CO data in order to check any systematic errors in the calibration of the $AKARI$ diffuse map, according a knowledge that the PAH distribution well represent the interstellar environment as well as interstellar molecular gas traced by CO \citep[e.g.,][]{2008ARA&A..46..289T}.

\section{Datasets}
\subsection{CO}
$^{12}$CO($J$=1--0) datasets were taken from the NANTEN Galactic Plane Survey \citep[NGPS;][]{2004ASPC..317...59M} with the NANTEN telescope. The NGPS observations were carried out from 1999 to 2003 with $\sim$1.1 million data points in total, covering $-$180$^\circ$--$60^\circ$ width in Galactic longitude and 10$^\circ$--20$^\circ$ width in Galactic latitude at angular resolution of 156$''$ and grid spacing of $4'$--$8'$.

\subsection{PAH}
We use the $AKARI$ 9 $\mu$m (6.7--11.6 $\mu$m band) preliminary datasets calibrated by the non-linearity correction, the reset anomaly correction, and the correction for the scattered light caused in camera optics (see more details in Amatsutsu et al., 2016 in prep.). The $AKARI$ data are smoothed with a Gaussian kernel to match the angular resolution and the grid size of the NGPS data.

\section{Region Selection}
We selected 10$^\circ$$\times$10$^\circ$ region near the Galactic center containing massive star-forming H{\sc ii} regions NGC 6334 and NGC 6357, suitable for checking the calibration accuracy where the PAH and CO emission is very strong.

NGC 6334 and NGC 6357 are thought to have similar distances and ages \citep[$\sim$2 kpc and $\sim$5 Myr, e.g.,][]{2012A&A...538A.142R,2007A&A...473..437A}, while NGC 6357 with more luminous stars including two O3-type stars has stronger 2.7 GHz radio continuum emission than NGC 6334 \citep{2012A&A...538A.142R}.

\begin{figure*}
\begin{center}
\includegraphics[width=140mm,clip]{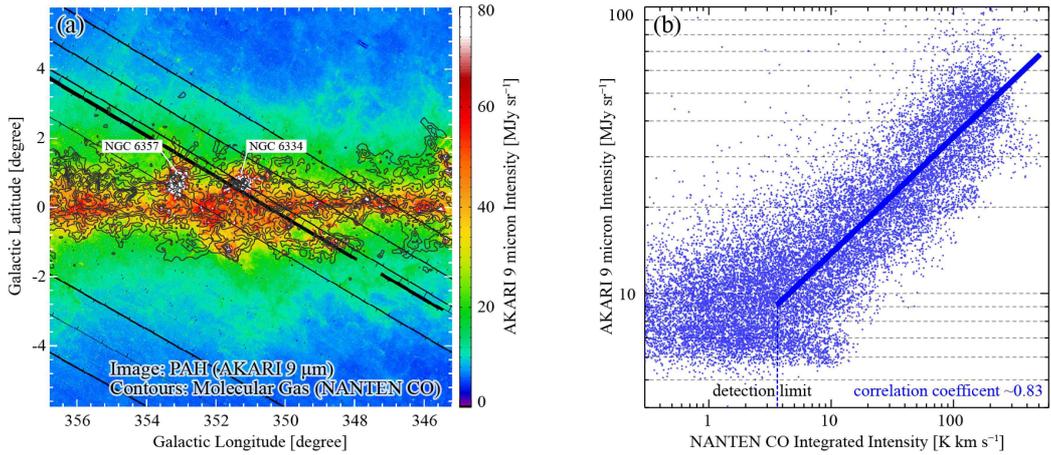}
\caption{(a) Distribution of the AKARI 9 $\mu$m intensity on which the NANTEN CO integrated intensity contours are overlaid \citep{2004ASPC..317...59M}. (b) Correlation diagram between the NANTEN CO integrated intensity (in unit of K km s$^{-1}$) and the AKARI 9 $\mu$m intensity (in unit of MJy sr$^{-1}$). The solid line indicates the best fitting curve with a power-law function.}
\label{fig1}
\end{center}
\vspace*{-0.15cm}
\end{figure*}%

\begin{figure}[h]
\begin{center}
\includegraphics[width=57mm,clip]{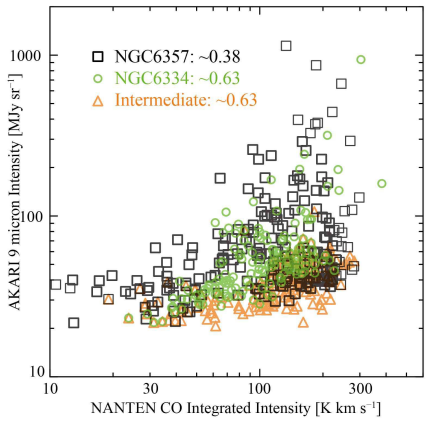}
\caption{Correlation diagram between the NANTEN CO integrated intensity and the AKARI 9 $\mu$m intensity toward NGC 6334, NGC6356, and between the two regions.}
\label{fig2}
\end{center}
\vspace*{-0.15cm}
\end{figure}%

\section{Results and Discussion}
Figure \ref{fig1}a shows the $AKARI$ 9 $\mu$m image superposed with the NANTEN CO contours. Figure \ref{fig1}a indicates that CO and PAH distributions show a good spatial correlation. Figure \ref{fig1}b shows a correlation plot between the CO and PAH intensities for the whole region in Figure \ref{fig1}a, showing a good correlation with a correlation coefficient of $\sim$0.83 above the CO detection limit. The relationship between integrated CO intensity $W$(CO) and 9 $\mu$m intensity $I_{\mathrm{9\mu m}}$ is well expressed by a power-law function as $I_{\mathrm{9\mu m}}$ = (5.4 $\pm$ 0.1) $\times$ $W$(CO)$^{(0.406 \pm 0.006)}$.

\begin{table}[h]
\caption{Correlation coefficient for each sampling grid\label{tab1}}
\centering
\begin{tabular}{lrrrr}
\toprule
 Sampling grid   &  4$'$  & 8$'$ & 16$'$ & 32$'$ \\
\midrule
Correlation coefficient & 0.83 & 0.83 & 0.82 & 0.84\\
\bottomrule
\end{tabular}
\vspace*{-0.15cm}
\end{table}

We also calculated the correlation coefficients changing sampling grid size to be 8$'$, 16$'$, and 32$'$. As a result, the correlation coefficients are nearly consistent for each grid size (see Table \ref{tab1}). Therefore, the calibration has no systematic error in the scale of 10$^\circ$ $\times$ 10$^\circ$.

Finally, we compare these correlations among three regions of NGC 6334, NGC 6357, and the intermediate region between the two (Figure \ref{fig2}). Each area of about 2.5 degree$^2$ is extracted. It reveals that the correlation coefficient of NGC 6357 ($\sim$0.38) is lower than the rest two regions ($\sim$0.63). We presume that this trend can be explained by the higher ionization degree and/or 
the PAH excitation in NGC 6357 by stronger UV radiation from local O-type stars. NGC 6357 is suggested to be more evolved indicated by the ratios of mid-infrared to far-infrared emission \citep{2012A&A...538A.142R}, consistent with the above idea. Further detailed analysis will enable us to understand the evolution and CO dissociation of NGC 6334 and NGC 6357.

\section{Summary}
We investigate the correlation of CO and PAH emission toward the massive star-forming regions NGC 6334 and NGC 6357 in order to evaluate the calibration accuracy of the $AKARI$ data. We confirmed that the $AKARI$ 9 $\mu$m intensity has a good positive correlation with the CO intensity in a scale of 10$^\circ$$\times$10$^\circ$, implying fairly good calibration of $AKARI$. We also suggest that the difference of ionization degree, PAH excitation, and evolutional stage between NGC 6334 and NGC 6357 results in the poorer correlation for small scale structures.


\acknowledgments
The NANTEN project was based on mutual agreements between Nagoya Univ. and the Carnegie Inst. of Washington. This research is based on the observations with $AKARI$, the JAXA project with the participation of ESA. This work is supported by the Grant-in-Aid for the Scientific Research Fund (No. 24740122, No. 26707008, and No. 50377925) from the Ministry of Education, Culture, Sports, Science and Technology of Japan.


\end{document}